\documentstyle[aps,preprint,float,prl]{revtex}
\tighten
\input{boxedeps.tex}
\SetRokickiEPSFSpecial
\HideDisplacementBoxes

\begin{document}
\draft

\title{Calculation of Higher Order Effects in Electron-Positron Pair 
Production in Relativistic Heavy Ion Collisions}
\author{Kai Hencken and Dirk Trautmann}
\address{
Institut f\"ur theoretische Physik der Universit\"at Basel,
Klingelbergstrasse 82, 4056 Basel, Switzerland
}
\author{Gerhard Baur}
\address{
Institut f\"ur Kernphysik (Theorie), Forschungszentrum J\"ulich, 
52425 J\"ulich, Germany
}

\date{August 13, 1998}

\maketitle

\begin{abstract}
We present a calculation of higher order effects for the impact
parameter dependent probability for single and multiple
electron-positron pairs in (peripheral) relativistic heavy ion
collisions. Also total cross sections are given for SPS and RHIC
energies. We make use of the expression derived recently by several
groups where the summation of all higher orders can be done
analytically in the high energy limit. An astonishing result is that
the cross section, that is, integrating over all impact parameters, is
found to be identical to the lowest-order Born result for symmetric
collisions. For the probability itself on the other hand we find
rather large effects at small impact parameters compared to the lowest
order results, which translate to large effects for the cross section
for multiple pair production.
\end{abstract}


Electron-Positron pair production in (peripheral) relativistic heavy
ion collisions has attracted some interest recently due to the
observation that at the relativistic heavy ion collider RHIC and LHC,
the probability for this process calculated in perturbation theory
violates unitarity, that is, gets larger than one, even for impact
parameters of the order of the Compton wave length $\lambdabar_C
\approx 386$~fm. This fact was first shown in
\cite{BertulaniB88,Baur90}. The unitarity violation was then studied
in a number of articles, taking into account higher order processes in
\cite{Baur90,RhoadesBrownW91,BestGS92,HenckenTB95,AlscherHTB97}.  It
was found that the inclusion of these higher-order processes leads to
the restoration of unitarity but also to new effects, mainly the
production of multiple pairs.  All studies also found that the
probability for $N$-pair production can be approximately described by
a Poisson distribution. Therefore the probability from perturbation
theory (in the following called ``reduced probability'') has to be
interpreted as the average number of pairs produced in one collision.
Deviations from the Poisson distribution where studied for small
impact parameters in \cite{HenckenTB95} and found to be rather small.

Calculations of the impact parameter dependent probability in the
external field approximation where calculated exactly in lowest order
for small impact parameters $b$ \cite{HenckenTB94} and later also for
all impact parameters \cite{GucluWUSE94,HenckenTB95b}. The total cross
section for pair production (that is, integrated over impact
parameters) was also calculated. Using a Monte Carlo approach it was
studied in \cite{BottcherS89}. An analytical form of the differential
cross section was found in \cite{AlscherHTB97}.  Cross sections for
multiple pair production were also given there.

One open question is still the importance of higher order corrections
--- or even nonperturbative effects --- coming from the large value of
the effective coupling constant $Z\alpha \approx 0.6$. Coupled channel
calculations have been done at smaller energies (up to $\gamma\approx
2$ in the collider frame). They always predicted a much larger
probability compared to perturbation theory
\cite{MombergerGS91,RumrichSG93}. But recently the accuracy of these
calculations has been questioned and by using a larger basis set
smaller results were found. In the end results were only a factor of
four larger than the perturbative results \cite{MombergerBS95}, which
is also in agreement with calculations using a spline-approach
\cite{ThielHGS95}.

In a recent article the summation of the effect of the target to all
orders was studied in the high energy limit (that is up to lowest
order in $1/\gamma$) for the related problem of bound-free pair
production \cite{Baltz97}. Here the electron is created into a bound
state of one of the ions. It was found that the summation to all
orders could be done analytically and a fairly simple expression was
found. The calculated probability for bound-free pair production was
found to be slightly smaller than the first order calculations.

A number of authors have in the mean time extended this approach also
to the calculation of (free) electron-positron pairs. In a first
article \cite{SegevW98} it was shown that here the summation can be
done analytically again, leading to a rather simple modification of
the matrix element. The authors of \cite{BaltzM98} come to the same
conclusion; they show also by integrating over the impact parameter
that the cross section becomes identical to the lowest order Born
result. In \cite{SegevW98b} the same conclusion is given. In
\cite{EichmannRSG98} the scattering of electrons in the field of
colliding nuclei is studied, a problem which is closely related to
pair-production.

\begin{figure}
\begin{center}
\ForceHeight{6cm}
\BoxedEPSF{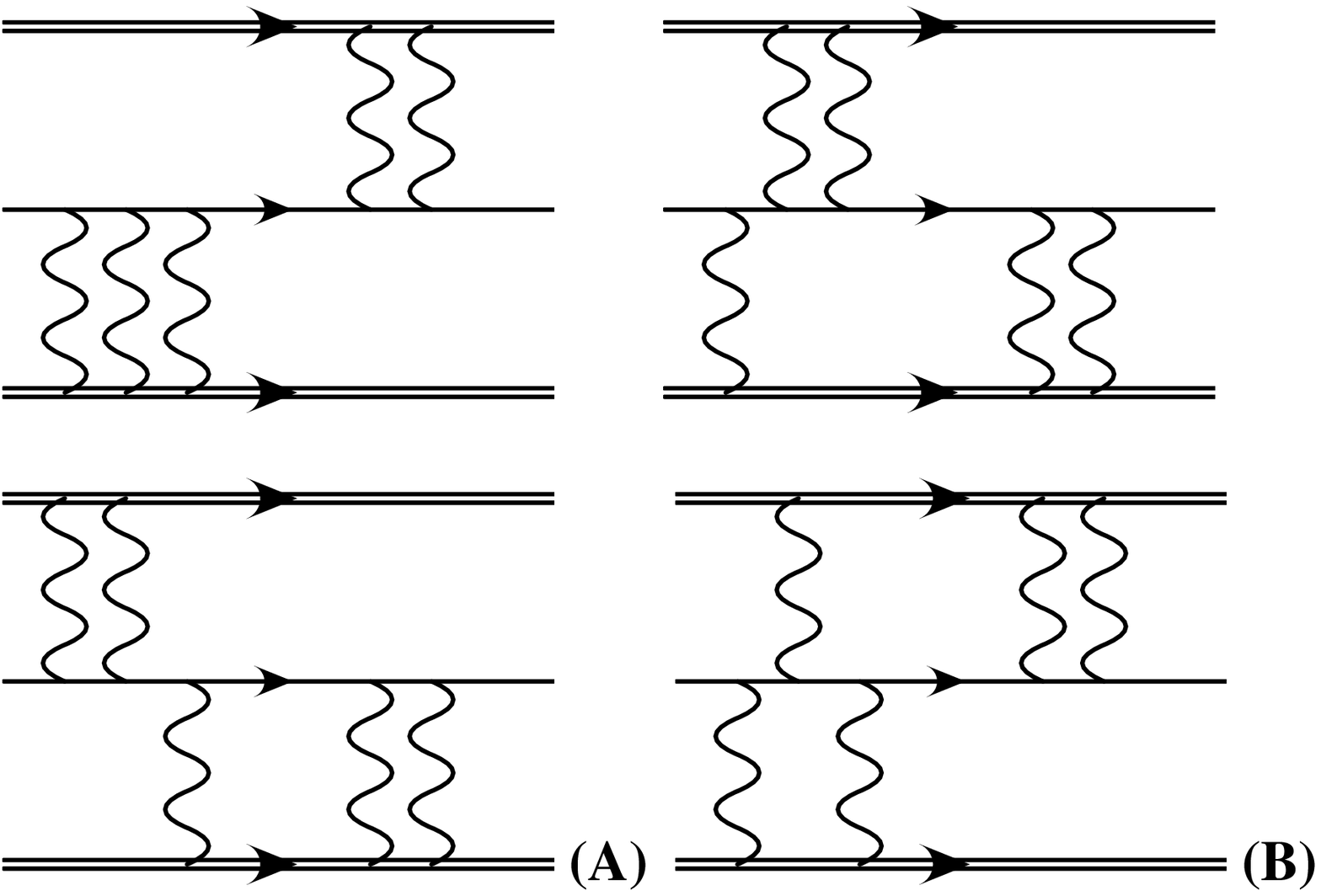}
\end{center}
\caption{Diagrams of the type (A) are included in the matrix
element. Diagrams of type (B) are assumed to be subdominant for large
$\gamma$.} 
\label{fig_feynman}
\end{figure}

Of course there remain a few questions that still need to be
addressed. The calculations were done using the Dirac-sea picture
(that is, starting with an electron with negative energy in the
initial state) and are therefore one-particle approximations of the
full problem. The applicability of this approximation for situations
where strong fields produce many particles has still to be
investigated. The use of the reduced probability in the Poisson distribution
to get multiple pair production was only derived in the (usual) Feynman
picture. Also only a part of all possible diagrams are included in the
approach. Fig.~\ref{fig_feynman}(A) shows typical diagrams which are
included, whereas those of Fig.~\ref{fig_feynman}(B) are not. In
\cite{EichmannRSG98} it is argued that this second class of diagrams
vanish in the limit $\gamma\rightarrow\infty$ for electron scattering,
therefore the neglect of them seems to be justified. But such a
rigorous proof is still needed for pair production. Finally it is well
known from calculations with the approach of Bethe, Maximom and Davis
that Coulomb corrections exist and are not small at the energies
considered \cite{BertulaniB88,IvanovKSG98}. This seems to be in
contrast to the observation that the cross section should be identical
to the lowest-order one for ion collisions. The most likely
explanation of their absence is, that they are of subleading order in
$1/\gamma$ and were therefore dropped.

It is our main aim to present a numerical calculation of the higher
order effects in these expressions given in these articles in order to
study their magnitude and their practical implications. Whereas the
cross section for one-pair production is dominated by the large impact
parameters, multiple-pair production is more sensitive to small impact
parameters. Therefore we try to see whether a measurement of multiple
pairs can be used to look for these higher order effects.

We make use of the expression Eq.~(54) of \cite{BaltzM98}. In
addition we also keep all effects of finite $\gamma$ in the
expression. Written in our notation the (reduced) differential
probability is
\begin{equation}
P(p_+,p_-,b) = \int d^2 \Delta q \tilde P(p_+,p_-,\Delta q) \exp(i
\Delta \vec q \vec b)
\end{equation}
where $p_+$ and $p_-$ are the momenta of positron and electron and the
Fourier-transform of the probability $\tilde P$ is given for symmetric
collisions ($Z=Z_A=Z_B$) by:
\begin{eqnarray}
&&\tilde P(p_+,p_-,\Delta q) = \frac{4\eta^4}{\beta^2} \int d^2q 
\nonumber\\
&&\biggl\{[-q^2]^{1+i\eta} [-q'^2]^{1-i\eta}
[-(q-(p_++p_-)^2)]^{1+i\eta} 
\nonumber\\
&&[-(q'-(p_++p_-)^2)]^{1-i\eta} \biggr\}^{-1}
\mbox{Tr} \biggl\{({\not\!p}_-+m) \nonumber\\
&&\left[
 \frac{{\not\!\!w}_1({\not\!p}_--{\not\!q+m}){\not\!\!w}_2}
{-(q-p_-)^2+m^2}
+\frac{{\not\!\!w}_2({\not\!q}-{\not\!p}_++m){\not\!\!w}_1}
{-(q-p_+)^2+m^2} 
\right]
({\not\!p}_+-m) \nonumber\\&&
\left[
 \frac{{\not\!\!w}_2({\not\!p}_--{\not\!q'}+m)
{\not\!\!w}_1}{-(q'-p_-)^2+m^2}
+\frac{{\not\!\!w}_1({\not\!q'}-{\not\!p}_++m)
{\not\!\!w}_2}{-(q'-p_+)^2+m^2}
\right]
\biggr\}, \nonumber\\&&
\end{eqnarray}
with $\eta=Z\alpha$, $q'=q+\Delta q$, $w_1=(1,0,0,\beta)$,
$w_2=(1,0,0,-\beta)$ and $m$ the electron mass. This expression is
almost identical to the one from perturbation theory, as given in
\cite{HenckenTB95b}; the only difference is the additional exponents
$1\pm i \eta$ for the photon propagator. This fact was already
observed in \cite{BaltzM98,SegevW98b}. Integrating over $b$ leads to a
delta-function $\delta(\Delta q)$. Then the photon propagators are
just complex conjugate to each other and only the absolute value
enters. Therefore the cross section calculations of
\cite{BottcherS89,AlscherHTB97} are exact in the high energy limit.

Here we want to study the effect of the higher orders for small impact
parameters. The total (single pair) cross section is completely
dominated by the large impact parameters, especially in the high
energy limit. Stronger deviations are to be expected mainly for small
$b$, especially if $b$ gets smaller than the Compton wavelength
$\lambdabar_C \approx 386 fm$.

The new expression has a complex exponent, which makes the expressions
oscillatory. Therefore a direct Monte Carlo integration is not
possible. We rewrite it into a form with only standard Feynman
integrals. We start by applying the usual Feynman-trick to group a
product of two denominator into a single one, integrating over an
auxiliary parameter. This trick is normally used for integer
exponents. But it is easy to see by looking at a derivation of this in
terms of $B$-functions (see, e.g., \cite{PeskinS95}), that the same
expression also hold for complex exponents. We use it here in the
following form:
\begin{eqnarray}
\frac{1}{C^{1+i\eta} D^{1-i\eta}} &=& \frac{1}{B(1+i\eta,1-i\eta)}
\int_0^1 \frac{w^{1+i\eta} (1-w)^{1-i\eta} dw} {[wC+(1-w)D]^2}
\nonumber\\
\label{eq_feynman}
\end{eqnarray}

Rewriting both photon propagators, we get two auxiliary integrations:
$w_A$ and $w_B$.  In the new denominator the factor $\pm i \eta$ just
cancel and the remaining expression is of the form of (the square of)
a propagator. The remaining two-dimensional integral over $q$ can then
be done in the same way as discussed in
\cite{HenckenTB94,HenckenTB95b,AlscherHTB97}. We integrate over all
final states of the electron and positron using VEGAS \cite{VEGAS78}.

Due to the oscillatory behavior of the numerator at the boundaries, a
direct numerical evaluation of the auxiliary integrals is not useful.
We therefore expand $\tilde P$ for each $\Delta q$ in terms of
polynomials of the form $[w_A (1-w_A)]^k [w_B (1-w_B)]^l$. Terms up to
$k,l=5$ have been included in a fit, but convergence is already found
with smaller exponents. The integrals over these polynomials can be
expressed now in terms of the $B$-functions. Our approach has the
advantage that the coefficients of the polynomial expansion are
independent of $\eta$, apart from the trivial $\eta^4$
dependence. Therefore results for arbitrary $\eta$ can be calculated
with no extra effort. Fourier-transforming this expression now with
respect to $\Delta q$ gives us $P(b)$. The total cross section, that
is, integrated over $b$, is given directly by $\sigma= (2\pi)^2 \tilde
P(\Delta q=0)$.

We have calculated $\tilde P(\Delta q)$ and $P(b)$ for both SPS (Pb-Pb
collisions at $\gamma=10$) and RHIC (Au-Au collisions at $\gamma=100$)
conditions. As a check of the correctness of our calculations, we can
compare them for $\eta\rightarrow 0$ with the perturbative results in
\cite{HenckenTB95b}, where a different approach was used. We get
perfect agreement between those two, giving us confidence in our
procedure.

\begin{figure}
\begin{center}
\ForceHeight{9cm}
\BoxedEPSF{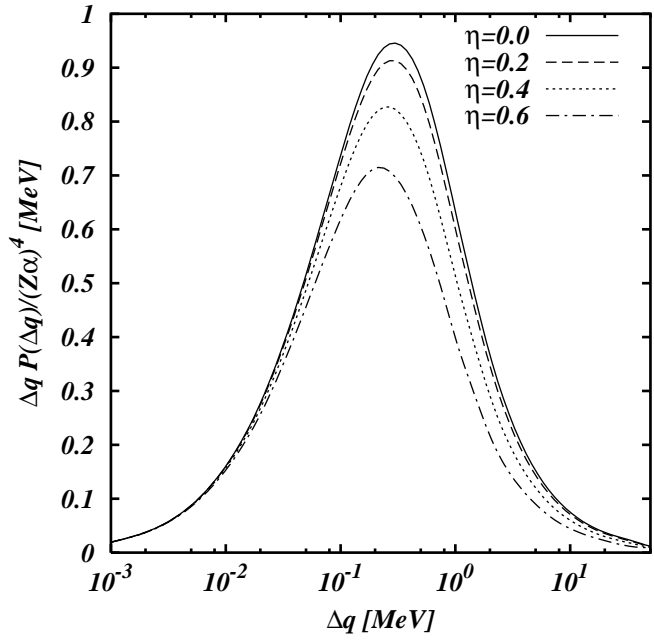}
\end{center}
\caption{$\Delta q P(\Delta q)$, the Fourier-transform of the total
probability is shown for $\gamma=100$. The trivial
$\eta^4=(Z\alpha)^4$ dependence is divided out. Shown are the results
for different $\eta$ for the full calculation.}
\label{fig_qpq100}
\end{figure}

Figure~\ref{fig_qpq100} shows the Fourier transform for $\gamma=100$
and for different $\eta$. The effect of the higher orders is quite
large, making $\tilde P$ smaller up to about 30\%. The shape of the
curve itself is changed only slightly. For $\Delta q \rightarrow 0$
all curves coincide with each other. This has to be the case as the
total cross section is identical to the lowest order one.

\begin{figure}
\begin{center}
\ForceHeight{9cm}
\BoxedEPSF{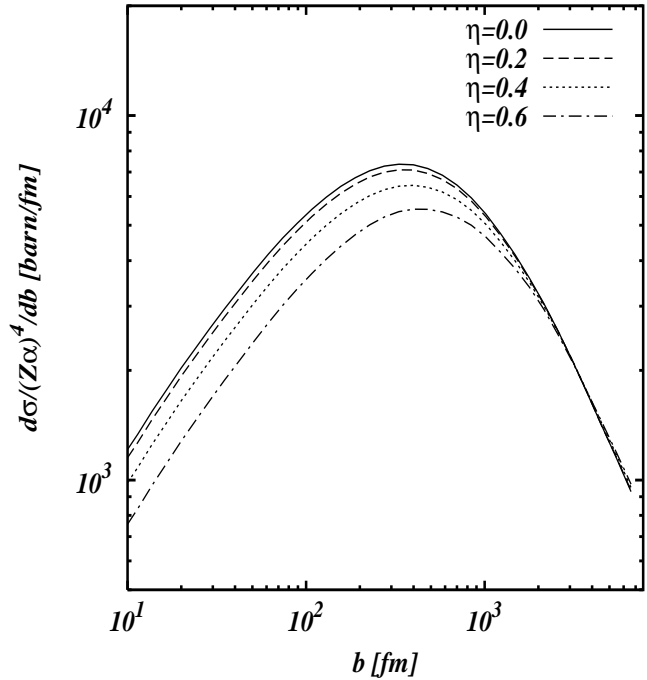}
\end{center}
\caption{Shown is the differential cross section $d\sigma/db$ for
$\gamma=100$ and for different values of $\eta$. The trivial
$\eta^4=(Z\alpha)^4$ dependence is divided out. At small impact
parameters the cross section is reduced quite substantially.}
\label{fig_dsigb100}
\end{figure}

Figure~\ref{fig_dsigb100} shows the impact parameter dependent cross
section $d\sigma/db=2\pi b P(b)$ for different values of $\eta$. A
deviation is only seen for small impact parameters, where it is quite
large.

\begin{figure}
\begin{center}
\ForceHeight{9cm}
\BoxedEPSF{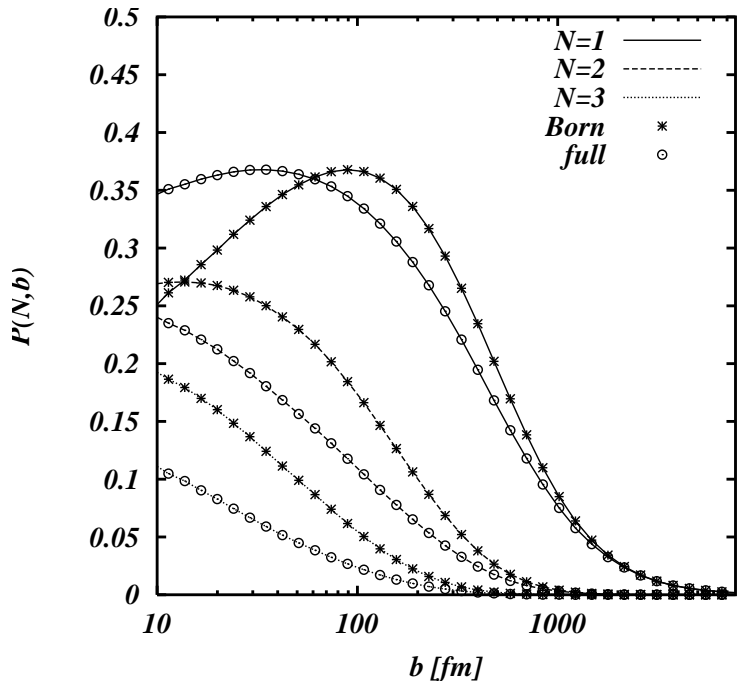}
\end{center}
\caption{The impact parameter dependent probability to produce $N$
pairs is shown for up to three pairs. Shown are results for the lowest
order Born result (stars) and also the full calculation (circles) both
for $\gamma=100$ and Au-Au collisions.}
\label{fig_pois100}
\end{figure}

Making now use of the Poisson distribution we can calculate
probabilities for multiple pair production.  For $\gamma=100$ we get
the results shown in Fig.~\ref{fig_pois100}. The higher order
processes reduce the multiple pair production probabilities, but the
probability for two-pair production is still large. Integrating over
$b$ we get the total cross sections as given in
table~\ref{tab_xsection}. For the single-pair cross section we use the
approach of \cite{AlscherHTB97}\footnote{This could be improved by
taking into account the Poisson distribution and also a lower bound
for the impact parameters. As these effects are rather small, we have
neglected them here.}. It is clearly seen that the cross section for
multiple pair production is sensitive to the higher order effects in
both cases. Therefore measuring them seems to be a practical way to
look for these effects.

In conclusion, we have calculated total probabilities for one and
multiple pair production including higher order effects using the
expression of \cite{BaltzM98}. This approach effectively sums up all
higher order diagrams in the high energy limit. We have
calculated cross sections for multiple pair production and have found
both probabilities and cross sections to be substantially smaller for
the full calculation compared to the perturbative one. Therefore a
measurement of this cross section should allow to really see these
higher order QED effects in an experiment.

Higher order effects can only be seen if one uses observables which
are sensitive to small impact parameters. Multiple pair production is
one such possibility. Depending on the experimental situation one
could also think of other ways, e.g., the production of other
particles together with an $e^+$-$e^-$ pair.

In this article we have concentrated on ``global properties'' of the
pair production, that is, total probabilities and cross sections. Our
main aim was to demonstrate that calculations are possible. In order
to see whether experiments will be able to see these effects, more
differential studies are needed. We will present these and also
details of the calculations in an upcoming publication.

\begin{table}[h]
\begin{center}
\begin{tabular}{ccc}
$N$      & Born (barn) & full (barn) \\
\hline
\multicolumn{3}{l}{$\gamma=10$, Pb-Pb ($Z=82$,$\eta=0.59$)}\\
1 & 4.21k & 4.21k \\
1 & 123 & 84.4 \\
2 & 8.61 & 3.88 \\
3 & 0.713 & 0.212 \\
\multicolumn{3}{l}{$\gamma=100$, Au-Au ($Z=79$,$\eta=0.57$)}\\
1 & 34k & 34k \\
2 & 893 & 624 \\
3 & 113 & 53.9 \\
4 & 18.9 & 6.04 \\
\end{tabular}
\end{center}
\caption{\sl Total cross section for one and multiple pair production
are given. Shown are the results for SPS/CERN and RHIC conditions. For
the results for $N=1$ the approach of \protect\cite{AlscherHTB97} was
used.} 
\label{tab_xsection}
\end{table}

\end{document}